\documentclass[aps,prl,twocolumn,english,superscriptaddress,nobibnotes,amsmath,amssymb,floatfix,longbibliography]{revtex4-2}

\usepackage[utf8]{inputenc}
\usepackage[T1]{fontenc}
\usepackage[english]{babel}
\usepackage{hyperref}
\usepackage{url}

\usepackage{changes}
\setcommentmarkup{\bfseries\color{red}-- #1 --}

\usepackage{amsmath}
\usepackage{amsfonts}
\usepackage{amssymb}
\usepackage{mathtools}
\usepackage{lmodern}
\usepackage{upgreek}
\usepackage{ragged2e}
\usepackage[version=3]{mhchem}
\usepackage[free-standing-units]{siunitx}

\usepackage{tabularx}
\usepackage{booktabs}
\usepackage{placeins,afterpage}
\usepackage{subcaption}
\DeclareCaptionLabelFormat{bold}{\textbf{(#2)}}
\DeclareCaptionJustification{justified}{\justifying}
\usepackage{epstopdf}
\usepackage{graphicx}

\usepackage{varwidth}
\usepackage{xcolor}
\definecolor{mossgreen}{rgb}{0.68, 0.87, 0.68}
\definecolor{mustard}{rgb}{1.0, 0.86, 0.35}
\definecolor{frenchblue}{rgb}{0.0, 0.45, 0.73}
\definecolor{lightskyblue}{rgb}{0.53, 0.81, 0.98}
\definecolor{purpleheart}{rgb}{0.41, 0.21, 0.61}
\definecolor{cadmiumred}{rgb}{0.89, 0.0, 0.13}

\usepackage{tikz}
\usetikzlibrary{calc,fit,shapes,arrows,positioning}
\usepackage{pgfkeys}
\usepackage{pgfplots}
\usepgfplotslibrary{fillbetween}
\tikzstyle{decision} = [diamond, draw, fill=cadmiumred!30, 
text width=4.5em, text badly centered, inner sep=0pt,outer sep = -.5pt]
\tikzset{block/.style = {
		rectangle, draw, fill=frenchblue!25, rounded corners, minimum height=3em,inner sep=5pt,%
		execute at begin node={\begin{varwidth}{#1}\centering},%
			execute at end node={\end{varwidth}}
	},
	block/.default={9em}
}
\tikzstyle{cloud} = [draw, ellipse,fill=mustard,
minimum height=2em]
\tikzstyle{line} = [draw, -stealth]

\usepackage{cleveref}

\crefmultiformat{figure}{figs.~#2#1\xdef\mycreffirstarg{#1}#3}%
{,#2{\crefstripprefix{\mycreffirstarg}{#1}}#3}{,#2#1#3}{ and~#2#1#3}
\usepackage{xparse}
\usepackage{xspace}

\ExplSyntaxOn
\prop_new:N \l_pollard_a_prop
\cs_new:Npn \pollard_wrappera:nn #1#2
{%
\subref*{#1}\nobreakspace#2\hspace{4pt plus 4pt minus 2pt}%
}

\NewDocumentCommand { \addsubcap } { m m }
{%
	{\phantomsubcaption\label{#1}}%
	\prop_put:Nnn \l_pollard_a_prop { #1 } { #2 }%
}
\NewDocumentCommand { \processCaptions } { }
{
	\prop_map_function:NN \l_pollard_a_prop \pollard_wrappera:nn
}
\ExplSyntaxOff

\newcommand{\abs}[1]{\left | #1 \right |}

\makeatletter 
\renewcommand{\fnum@figure}{\textbf{Figure~\thefigure}}
\makeatother

\def \Aeff{A_{\text{eff}}}
\def \At{\tilde{A}}

\begin{document}
\title{Dynamic Interplay Between Kerr Combs and Brillouin Lasing in Fiber Cavities}

\def \ICB{Laboratoire Interdisciplinaire Carnot de Bourgogne (ICB), UMR6303 CNRS-UBFC, 21078 Dijon, France}

\author{Erwan~Lucas}
\email[]{erwan.lucas@u-bourgogne.fr}
\affiliation{\ICB}

\author{Moise~Deroh}
\affiliation{\ICB}

\author{Bertrand~Kibler}
\affiliation{\ICB}

\begin{abstract}
	We study a re-configurable nonreciprocal fiber ring cavity setup to generate tunable optical frequency combs.
	Coherent combs can be obtained by finely exploiting bi-chromatic Brillouin lasing and cascaded four-wave mixing in the nonlinear fiber cavity.
	We introduce a  numerical model of the cavity system, which is validated by comparisons with experiments.
	Our work shows the importance of the mode pulling effect, controlled by the pump laser detuning, in setting the dynamics of the comb.
	This effect along with the impact of multimode lasing had been overlooked in previous fiber experiments.
	We discuss these limitations and devise several scaling laws for these system.
\end{abstract}

\maketitle
Optical frequency combs (OFCs), whose broadband spectrum is composed of equi-spaced coherent frequency tones, are very powerful tools in a wide range of fields including frequency metrology, microwaves generation, spectroscopy, and astronomical spectrograph calibration~\cite{Diddams2020,Picque2019}.
Various technologies can be used to generate such OFCs, such as mode-locked lasers, electro-optical modulation, and nonlinear frequency conversion in passive optical resonators, known as Kerr combs~\cite{Diddams2020}.
In this contribution, we study a hybrid system based on a bi-chromatic Brillouin fiber laser~\cite{Li2017c}, to generate a Kerr comb~\cite{Kippenberg2018}.
This scheme has been recently experimentally demonstrated in fiber cavities~\cite{Li2017c,Huang2019a} and microresonators~\cite{Bai2020,Do2021}.
However, a detailed analysis of the dynamical interplay between the Brillouin lasing and the Kerr comb generation remains elusive, especially in non-reciprocal fibers cavities.
Here, we investigate these systems in depth, accounting for the complex nature of the Brillouin gain, and highlight the role of the pump detuning parameter.
Most importantly, we observe and discriminate stable single-frequency Brillouin lasing regimes yielding low-noise Kerr combs, in contrast with noisy regimes, where lasing occurs over multiple cavity modes.
\begin{figure}[!tb]
	\centering%
	\begin{tikzpicture}[inner sep=1pt,font=\sffamily\footnotesize]
		\node (fig1) {\includegraphics[width=\linewidth]{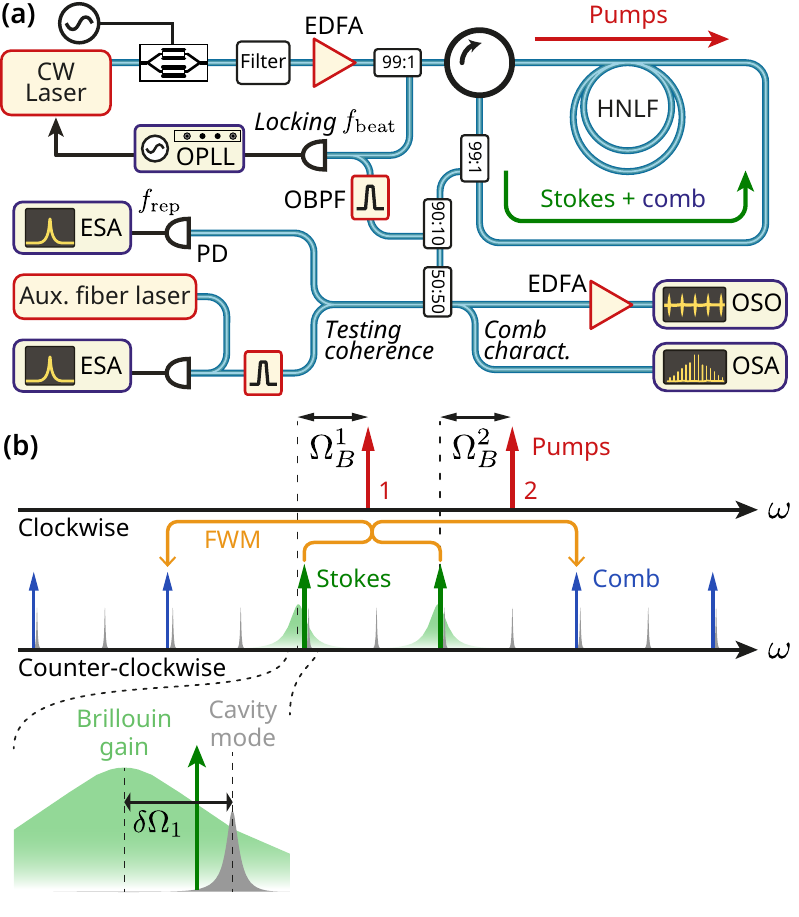}};
		\begin{axis}[
			at={(fig1.south east)},
			anchor=below south east,
			yshift=-1ex,
			name=areaPlot,
			scale only axis,
			width=.5\linewidth, height=.25\linewidth,
			enlargelimits=false, ymax=2,
			ticks=none,
			axis on top,
			xlabel near ticks, ylabel near ticks,
			xlabel style={yshift=-1mm}, ylabel style={yshift=1mm,align=center}, 
			xlabel={$D_1/\Gamma_B \propto 1/M$},
			ylabel={Pump power\\(arb. unit)}
			]
			\addplot[name path=thres,domain=0.2:1] {x}
			node [pos=1, above, sloped,color=blue,anchor=south east] {\begin{varwidth}{3cm}\raggedleft Coherent,\\single frequency\end{varwidth}};
			\addplot[name path=f,domain=0.2:1, color=red] {x*(1+2*x^2)};
			\path[name path=axisTop] (current axis.north west) -- (current axis.north east);
			\addplot[
			thick,
			fill=blue!20, 
			]
			fill between[
			of=thres and f,
			soft clip={domain=0:1},
			];
			\addplot[
			thick,
			color=blue,
			fill=red!15
			]
			fill between[
			of=axisTop and f,
			soft clip={domain=0:1},
			];
			\node[color=red] at (axis cs:.42,1.3) {Noisy, multimode};
			\node[above left = 2pt of current axis.south east] {Below threshold};
		\end{axis}
		\node[anchor=east, yshift=-1.5ex,fill=white, fill opacity=.8, text opacity=1, inner sep=3pt] at (areaPlot.above north east) {Shorter $ \rightarrow $};
		\node[anchor=west, yshift=-1.5ex,fill=white, fill opacity=.8, text opacity=1, inner sep=3pt] at (areaPlot.above north west) {$ \leftarrow $ Longer cavity};
		\node[anchor=west,inner sep=0] at (areaPlot.left of north west) {\footnotesize\textsf{\textbf{(c)}}};
	\end{tikzpicture}%
	\addsubcap{fig:Concept:Setup}{Schematic of the experimental setup. EDFA: Erbium-doped fiber amplifier, HNLF: Highly nonlinear fiber, OPLL: Optical phase lock loop, OBPF: Optical bandpass filter, ESA: Electronic spectrum analyzer, OSA: Optical spectrum analyzer, OSO: Optical sampling oscilloscope}%
	\addsubcap{fig:Concept:Scheme}{Principle of Brillouin-Kerr comb generation.}%
	\addsubcap{fig:Concept:NoiseLength}{Simplified stability chart of the lasing states, as a function of the cavity FSR relative to the Brillouin gain bandwidth, and of the pump power.}%
	\caption{\textbf{Kerr comb in fiber cavity Brillouin laser}.
		\processCaptions
	}
	\label{fig:Concept}
\end{figure}

Figure~\ref{fig:Concept:Setup} presents the experimental setup and its operating principle.
It is based on a typical Brillouin laser configuration~\cite{Shirazi2007,Danion2016}, where the fiber cavity is closed by an optical circulator.
A highly nonlinear fiber (HNLF) is used to enhance both Brillouin and Kerr nonlinearities.
Hence, the two pump lasers are not subject to any resonance condition in the clockwise direction and circulate over a single roundtrip.
This allows efficient injection of the pump lasers, while relaxing their linewidth requirement, as fiber cavities can exhibit ultra-high quality factors.
These pumps induce Brillouin Stokes gain in the counter-clockwise direction, which is down-shifted by the phonon frequency $ \Omega_B/2\pi $.
When the gain overcomes the loss, Stokes waves can resonate and build up to form laser lines, which in turn produce a comb via cascaded four-wave mixing (FWM).
As an added benefit, Brillouin Stokes lasers can feature a linewidth narrowing relative to their pump~\cite{Gundavarapu2019,Debut2000}, potentially enhancing the comb coherence.

For the sake of simplicity, we only consider the case of net normal cavity dispersion, where comb formation via dual pumping is possible~\cite{Hansson2014a}.
This allows us to avoid modulation instability, occurring in the anomalous dispersion regime, whose chaotic nature could interfere with our discussion on the impact of laser dynamics.
\noindent\emph{Model:}
We developed a specific model to simulate comb generation in this type of laser cavity, based on a set of coupled equations derived from the Ikeda map~\cite{Ikeda1979}.
At each roundtrip, two successive steps are performed.
The first one determines the steady-state Brillouin gain, as a function of the re-circulating Stokes power and input pump power.
The second implements the nonlinear wave propagation in the fibers in the counter-clockwise direction.
First, the Brillouin scattering \emph{steady state} is computed by solving the equations
\begin{align}
	\dfrac{d \abs{A^j}^2}{dz} &= \left( - \dfrac{\operatorname{Re}\bigl[ g_B(\delta\Omega_j) \bigr] }{\Aeff} \, \abs{\At_{\mu_j}}^2 - \alpha \right) \, \abs{\A^j}^2
	\\
	- \dfrac{d \abs{\At_{\mu_j}}^2}{dz} &= \left( \dfrac{\operatorname{Re}\bigl[ g_B(\delta\Omega_j) \bigr]  }{\Aeff} \, \abs{A^j}^2 - \alpha \right) \, \abs{\At_{\mu_j}}^2
	\\
	\left| A^j(L) \right|^2 &= P^p_{\text{in}}
	\qquad \text{and} \qquad
	\At_{\mu_j}^{(n)}(0) = \At_{\mu_j}^{(n-1)}(L)
	\\
	g_B(\delta\Omega) &= \dfrac{g_B^0}{1+2i \, \delta\Omega/\Gamma_B}
\end{align}
where $ \abs{A^j}^2 $ and  $ \abs{\At_{\mu_j}}^2 $ are the powers of the pump and corresponding Stokes wave. $ \mu_j \approx \Omega_B/D_1 $ is the relative mode number of the cavity experiencing Stokes lasing, and $ D_1/2\pi $ is the cavity's free spectral range (FSR).
The index $ j=1,2 $ accounts for each pump.
$ \alpha $  is the attenuation coefficient in the fiber, which can usually be neglected compared to the magnitude of the discrete losses in the cavity (circulator, coupler, splices, \dots).
$ \Gamma_B $ is the Brillouin gain bandwidth, while $ \delta\Omega $ is the mismatch between the Brillouin gain peak and the nearest lasing mode, which can be controlled by tuning the pump laser frequency.
We solve this step numerically using a boundary value problem solver~\cite{Kierzenka2001}, which provides the spatial distribution of the pump power $ A^j(z) $ for the next step.

In the second step, the following modified Nonlinear Schrödinger Equation (NLSE), is integrated using the split-step Fourier method:
\begin{equation}\label{NLSEB}
	\begin{aligned}
		\frac{\partial \At_\mu}{\partial z} &=
		\left[
		-\frac{\alpha}{2}
		- i\frac{\beta_2}{2} \omega_\mu^2
		\right] \At_\mu 
		+ i \gamma_0 \ \mathcal{F} \left[ \left|A\right|^2 A \right]_\mu \\
		& + \sum_{j=1,2} \dfrac{g_B(\delta\Omega_j)}{2 \, \Aeff} \abs{A^j(L-z)}^2  \At_\mu  \, \delta_{\mu_j} 
	\end{aligned}
\end{equation}
The equation is written in the frequency domain, $ \mu $ represents the longitudinal mode number in the cavity. The Brillouin gain term is applied to the pumped Stokes modes $ \mu_{j},\ j\in{1,2} $ ($\delta_{ \mu_j} $ represents the Kronecker symbol). The operator $ \mathcal{F} \left[ \cdot \right]_{\mu} $ represents the $ \mu^{\text{th}} $ component of the Fourier series.

Finally, at the end of the propagation, the out-coupling (ratio $ \theta $) and discrete losses ($ \eta $) are applied to iterate from roundtrip $ n \rightarrow n+1 $: 
$ \At_\mu^{(n+1)}(0) = \sqrt{\eta \, \left( 1-\theta \right)} \, \At_\mu^{(n)}(L)  $
\noindent\emph{Experimental validation:}
We studied two cavities (A,B), whose attributes are summarized in \cref{table:parameters}.
They differ mainly on the HNLF length, thus impacting the resulting comb dynamics.
We minimized the amount of standard SMF fiber introduced by the coupling components, to maintain a normal net dispersion.

\begin{table}
	\begin{tabularx}{\linewidth}{@{\;}XSS@{\quad}l}
		\multicolumn{1}{r}{Cavity} & \multicolumn{1}{c}{A} & \multicolumn{1}{c}{B} & \multicolumn{1}{c}{Unit} \\
		\toprule
		\multicolumn{4}{l}{\textbf{HNLF}} \\
		\midrule
		Length $ L $ &  10 & 60 & \si{\meter} \\
		Brillouin gain $ g_B^0/A_{\text{eff}} $ &  3 & 3 & \si{\decibel\per\meter\per\watt} \\
		Gain bandwidth $ \Gamma_B /2\pi $ & 52 & 47 & \si{\mega\hertz} \\
		Nonlinearity $ \gamma $ &  12.5 & 9 & \si{\per\watt\per\kilo\meter}\\
		GVD $ D $ &  -2.9 & -0.55 & \si{\pico\second\per\nano\meter\per\kilo\meter} \\
		\addlinespace[1ex]
		\multicolumn{4}{@{}l}{\textbf{Other cavity parameters}}\\
		\midrule
		Length SMF28 $ L_{\textsc{smf}} $ &  1.2 & 5 & \si{\meter} \\
		Loss $ \eta $ & -1.17 & -1.22 & \si{\decibel\per\text{roundtrip}}  \\
		Out-coupler $ \theta $ &  1 & 1 & \si{\percent} \\
		Loaded linewidth $ \kappa/2\pi $ & 695 & 128 & \si{\kilo\hertz} \\
		Free spectral range $ D_1/2\pi $ & 18.39 & 3.15 & \si{\mega\hertz}
	\end{tabularx}
	\caption{Physical parameters of the cavities studied in this work. GVD: Group-velocity dispersion.}
	\label{table:parameters}
\end{table}
\begin{figure*}
	\centering%
	\includegraphics[width=\linewidth]{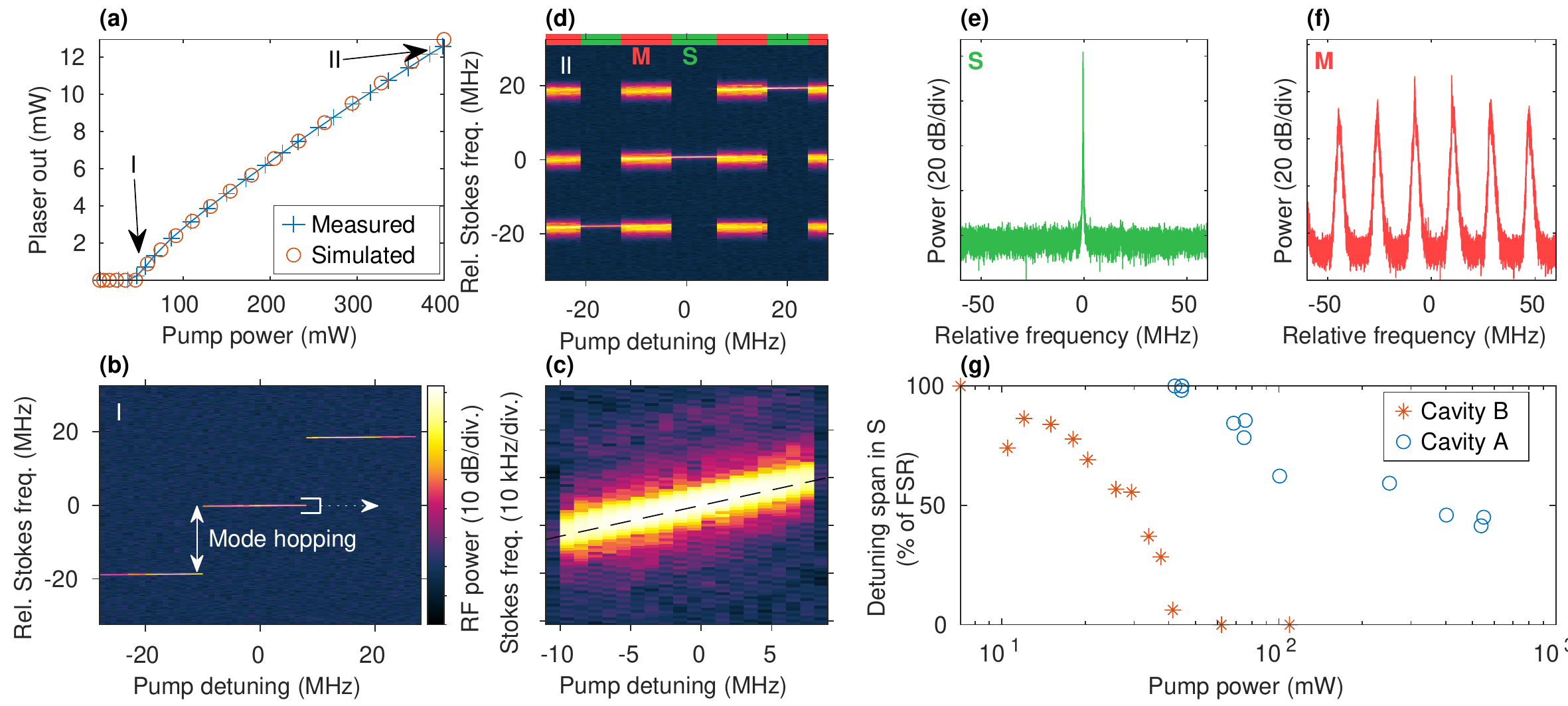}%
	\addsubcap{fig:BrillouinLase:Threshold}{Lasing threshold measurement in cavity A and simulation.}%
	\addsubcap{fig:BrillouinLase:modeHopLowP}{Spectrogram of the Stokes laser spectrum with the pump frequency, near threshold (I).}%
	\addsubcap{fig:BrillouinLase:modeHopzoom}{Zoom in on the step in \subref{fig:BrillouinLase:modeHopLowP}, showing the mode pulling effect.}%
	\addsubcap{fig:BrillouinLase:modeHopHighP}{Evolution of the Stokes laser spectrum with pump detuning at higher power (II). The laser is multimode around mode-hopping regions (red, \textsf{M}), while the detuning range with single mode emission (green, \textsf{S}) shrinks with increased power.}%
	\addsubcap{fig:BrillouinLase:Smode}{Stokes laser -- pump laser beatnote in the single mode lasing regime.}%
	\addsubcap{fig:BrillouinLase:Mmode}{Same beatnote in the multimode lasing regime.}%
	\addsubcap{fig:BrillouinLase:DetuningRange}{Evolution of the pump detuning range for single frequency emission (\textsf{S}) with the pump power for the two different cavity lengths.}%
	\caption{%
		\textbf{Brillouin laser mode pulling and emission regimes characterization}. \processCaptions
	}
	\label{fig:BrillouinLase}
\end{figure*}
\begin{figure*}[t]
	\centering%
	\includegraphics[width=\linewidth]{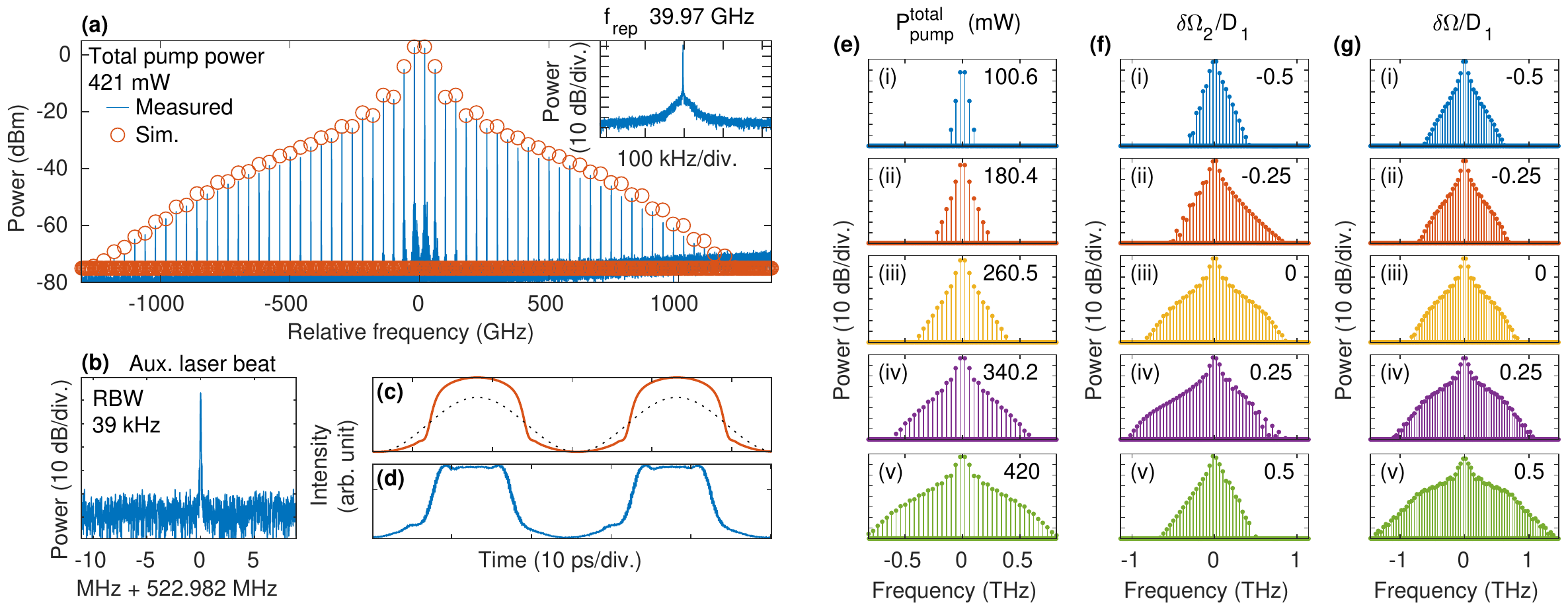}%
	\addsubcap{fig:CombQuiet:Comb}{Optical spectrum of a comb obtained in cavity A, with \SI{40}{\GHz} pumps separation. The simulation result is overlaid (circles). The pump power is \SI{421}{\milli\watt}. The inset shows the repetition rate beatnote of the comb. The resolution bandwidth (RBW) is \SI{100}{\hertz}.}%
	\addsubcap{fig:CombQuiet:auxbeat}{RF beatnote of the comb line $ -10 $ with a reference auxiliary laser.}%
	\addsubcap{fig:CombQuiet:temporalProfilesSim}{Simulated intra-cavity temporal intensity profile (solid), and pump beat pattern (dotted).}%
	\addsubcap{fig:CombQuiet:temporalProfilesMeas}{Temporal intensity profile corresponding to the comb shown in \subref{fig:CombQuiet:Comb}, measured on the OSO.}%
	\addsubcap{fig:CombQuiet:simPowScan}{Simulated pump power sweep from  \SIrange{100}{420}{\milli\watt} for $ \delta\Omega_1 = \delta\Omega_2 = 0 $.}%
	\addsubcap{fig:CombQuiet:simdesyncScan}{Simulated scan of $ \delta\Omega_2 $ from $ -D_1/2 $ to $ D_1/2 $, while $ \delta\Omega_1 = 0 $ is kept constant. The pump power is \SI{420}{\milli\watt}.}%
	\addsubcap{fig:CombQuiet:simDetScan}{Simulated detuning scan of $ \delta\Omega = \delta\Omega_1 = \delta\Omega_2 $ from $ -D_1/2 $ to $ D_1/2 $ for a pump power of \SI{420}{\milli\watt}.}%
	\caption{\textbf{Comb generation in single mode regime and comb evolution}.
		\processCaptions
	}
	\label{fig:CombQuiet}
\end{figure*}

First, we characterize the Brillouin laser with a single pump, thus disabling the Kerr comb generation.
The threshold was measured at \SI{42}{\milli\watt} in the short cavity A (see \cref{fig:BrillouinLase:Threshold}), while it was notably smaller (\SI{7.5}{\milli\watt}) in the longer cavity B.
The values of $ g_B/\Aeff \sim \SI{3}{\decibel\per\watt\per\meter} $ and Stokes shift of $ \Omega_B/2\pi \sim \SI{9.1}{\giga\hertz} $ were obtained in both fibers, in agreement with typical HNLF characteristics~\cite{DerohLargeBrillouin2018}.

The Stokes shift was measured by beating the Stokes laser with the pump laser.
Monitoring this beatnote frequency $ f_{\text{beat}} $, while scanning the Brillouin gain detuning $ \delta\Omega $, allows us to precisely track the emission frequency of the Brillouin laser.
We implemented an offset sideband locking~\cite{Thorpe2008b}, of the pump to the cavity, such that $ \delta\Omega $ is controlled by the synthesizer used to generate the sideband ($ f_{\text{synth}} $).
Thus the variation of the Stokes emission frequency can be expressed as $ \Delta f_{\text{Stokes}} = \xi \Delta f_{\text{synth}}  - \Delta f_{\text{beat}} $, where $ \xi = - 1 $ since the lower frequency sideband was selected.

\Cref{fig:BrillouinLase:modeHopLowP} shows the laser frequency tuning in cavity A near the lasing threshold.
Mode hopping of the Stokes laser is clearly visible as the emission frequency switches from a single longitudinal mode to another, providing a good estimate of the cavity FSR.
Focusing on range sustained by a single longitudinal mode, we see that the emission frequency shifts linearly as a function of $ \delta\Omega $ (\cref{fig:BrillouinLase:modeHopzoom}).
This phenomenon is known as mode pulling and stems from the imaginary part of the Brillouin gain, in agreement with the Kramers-Kronig relations~\cite{Nicati1994a,Li2012a,Do2021}.
As a result, the laser emission is steered toward the peak of gain lobe.
The mode pulling slope is fitted and follows the expression~\cite{Lecoeuche1996} $ \dfrac{\partial f_{\text{Stokes}}}{\partial\delta\Omega} = \dfrac{1}{1+\Gamma_B/\kappa} = \num{1.33e-2} $, yielding the ratio of the Brillouin gain bandwidth to the cavity linewidth $ \Gamma_B/\kappa = 74 $.
As $ \kappa $ was measured, we retrieve $ \Gamma_B/2\pi = \SI{52}{\mega\hertz} $ and a similar value for cavity B, which is typical for HNLF~\cite{DerohLargeBrillouin2018}.

At higher power (\cref{fig:BrillouinLase:modeHopHighP}), the laser dynamics changes significantly around the mode hopping transitions.
The emission goes from single-mode at small detuning (\textsf{S} regime, shown in \cref{fig:BrillouinLase:Smode}), to multimode regime at higher detuning, where the energy is distributed on several modes and the laser coherence is strongly degraded due to mode competition (\textsf{M} regime in \cref{fig:BrillouinLase:Mmode}).
Importantly, the detuning range where the \textsf{S}-mode is maintained, gradually shrinks with increased pump power, as shown in \cref{fig:BrillouinLase:DetuningRange}.
Beyond a critical power, the laser operates purely in the \textsf{M} regime.
Moreover, the loss of stability occurs at lower power in a long cavity than in a short one.
Indeed, in a long cavity, the number of modes under the gain curve is larger and their respective gains are similar, leading to simultaneous oscillations.

\noindent\emph{Comb generation:}
Bi-chromatic pumping can be achieved by combining two continuous-wave lasers or, as shown in \cref{fig:Concept:Setup}, by means of electro-optical modulation.
We use a Mach-Zehnder intensity modulator in carrier-suppressed mode driven by a tunable signal from a synthesizer (at \SI{20}{\giga\hertz}) to create two pumps with a tunable spacing $ \Delta\omega_p/2\pi \approx \SI{40}{\giga\hertz}$ (much higher than the cavity's FSR).
We use a programmable filter after the modulator to select only the first order sidebands and further reject the carrier.
To suppress the impact of the thermal drift of our cavity relative to the pump laser, we implement a phase-lock loop to frequency-lock the beatnote $ f_{\rm beat} $ of one of the pump with its Stokes laser~\cite{Danion2016}.

Bichromatic pumping presents the interesting feature of being thresholdless~\cite{Strekalov2009}.
Initial parametric sidebands are observed when the Stokes lasing threshold is reached for both Stokes waves.
The number of generated lines then increases with the injected pumps power, as shown in \cref{fig:CombQuiet:simPowScan}.
\Cref{fig:CombQuiet:Comb} shows a comb generated in cavity A, for a total pump power of \SI{421}{\milli\watt}.
The pump spacing and lockpoint were chosen so as to preserve a coherent singlemode emission and maximize the comb bandwidth.
The comb coherence is assessed by measuring a narrow RF beatnote for the comb repetition rate as well as for beat between the $ -10 $ comb line and an auxiliary low noise laser. (\cref{fig:CombQuiet:auxbeat}).
It is worth noting that the comb's repetition rate differs from the pump spacing.
Since the phonon frequency is wavelength-dependent, the Stokes 1 and 2 experience a slightly different shift from their respective pump:
$ \Omega_B^j = \dfrac{n_jv_A}{\lambda_j} $,
where $ n_j $ is the effective refractive index at the pump wavelength $ \lambda_j $ and $ v_A $ is the acoustic velocity.
This effect amounts to $ \Omega_B^2 - \Omega_B^1 \sim \SI{2}{\mega\hertz} $ in the present configuration.

Under these conditions, and by setting small pumps detunings in our model ($ \delta\Omega_1 = \delta\Omega_2 = \SI{-2}{\mega\hertz} $), the simulated comb agrees remarkably well with the experimental data.
Under normal net cavity dispersion, the corresponding temporal pattern in the cavity consists of a switching waves~\cite{Garbin2017} formed on the sinusoidal modulation pattern resulting from the interference between the two Stokes fields, as shown in \cref{fig:CombQuiet:temporalProfilesSim}.
The temporal waveform of the comb measured on an optical sampling oscilloscope (OSO) (\cref{fig:CombQuiet:temporalProfilesMeas}) is in good agreement with this prediction and further validates the comb coherence.
The deviation of the pulse shape is likely caused by propagation in the optical amplifier before the OSO.

Our simulations highlight the importance of the gain detuning parameters $ \delta\Omega_j $, even if the pumps are non-resonant.
The mode pulling effect induces a change of the emission frequency of the Stokes lasers and hence allows detuning control in the comb generation, which is a key parameter in setting the comb's dynamics~\cite{Lucas2017}.
First, an asymmetry in the detunings $ \delta\Omega_{1,2} $ has a significant impact on the spectral width and symmetry of the comb, as shown in \cref{fig:CombQuiet:simdesyncScan}.
The frequency spacing between the pumps $ \Delta\omega_p $ should be kept commensurate with the cavity's FSR, to maximize the comb bandwidth, similar to pulse-pumping Kerr comb generation~\cite{Obrzud2017}.
Second, even if $ \Delta\omega_p $ is an exact multiple of the FSR, increasing both detunings $ \delta\Omega_{1,2} $ toward the red helps to further expand the comb, as illustrated on \cref{fig:CombQuiet:simDetScan}.
	\begin{figure*}
		\centering%
		\includegraphics[width=\linewidth]{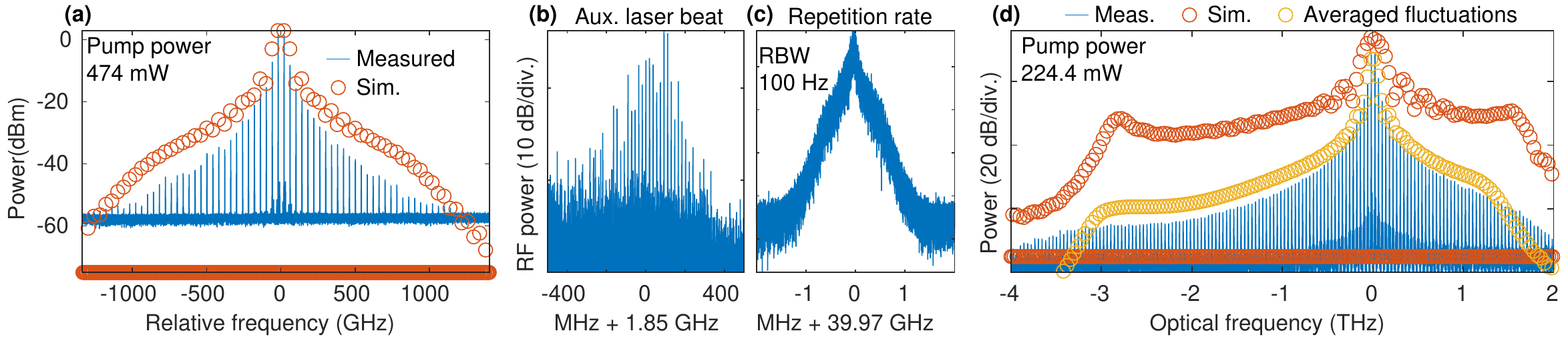}%
		\addsubcap{fig:CombNoisy:CombA}{Optical spectrum of a comb in the multimode Brillouin lasing regime (cavity A). The pump power is \SI{474}{\milli\watt}. The simulations are performed in the steady state and do not account for the instabilities, leading to a poor agreement.}%
		\addsubcap{fig:CombNoisy:auxbeat}{RF beatnote of the comb line $ -10 $ with a reference auxiliary laser (RBW \SI{39}{\kilo\hertz}). The multiple peaks result from the multi-mode operation.}%
		\addsubcap{fig:CombNoisy:frep}{Comb's repetition rate beatnote, evidencing the low coherence.}%
		\addsubcap{fig:CombNoisy:CombB}{Optical spectrum of a comb in cavity B in the multimode regime, for a pump power of \SI{224.4}{\milli\watt}.  The asymmetric combs shape is due to the third order dispersion. The red dots mark the simulated comb with static parameters ($ \delta\Omega_j = 0 $). The yellow dots are obtained by averaging 1120 spectra, calculated while the power and detuning fluctuate randomly in the range $ [0, 225] $~mW and $ [-D_1, D_1] $ respectively.}%
		\caption{\textbf{Multimode lasing and comb noise}.
			\processCaptions
		}
		\label{fig:CombNoisy}
	\end{figure*}%

\noindent\emph{Multi-mode lasing and limitations:}
When the Stokes lasers operate in the multimode regime, the comb coherence is strongly degraded, as shown in \cref{fig:CombNoisy}.
Such state features a rapid decrease in the power per line, and a significantly higher noise floor on the optical spectrum analyzer.
The measured beatnotes are considerably broader and multi-peaked.
The simulation -- experiment agreement also deteriorates (see \cref{fig:CombNoisy:CombA}), as the chaotic behavior of the Stokes lasers is not accounted for in our model.
Adding fluctuations in pump power and detuning into our model, and then averaging the resulting combs, provides a closer match to the experiments (see \cref{fig:CombNoisy:CombB}).

Mode hopping and multimode lasing are the two main sources of limitations in coherent Brillouin-Kerr comb generation.
First, these effects limit the detuning range for coherent comb generation, thus restricting the accessible comb states within the Brillouin laser.
For example, states (i) and (v) shown in \cref{fig:CombQuiet:simdesyncScan,fig:CombQuiet:simDetScan} are unrealistic in practice, since they require extreme detunings $ \delta\Omega_{1,2} = \pm D_1/2 $, where mode hop occurs.
Moreover, as the pump power increases, the detuning range for coherent comb generation further shrinks, as shown in \cref{fig:BrillouinLase:DetuningRange}.
Beyond a power level $ P_{\text{max}} $, both Stokes lasers operate only in multimode emission.
This maximum power depends on a few factors~\cite{Randoux1995,Lecoeuche1996}, the most prominent being the cavity's FSR compared to the Brillouin gain bandwidth.
The factor $ M = \Gamma_B/D_1 \propto L_{\text{tot}} $ relates to the number of modes under the gain curve and scales with the cavity length $ L_{\text{tot}} $.
The transition to multimode operation of the Stokes laser, due to optical mode competition over a common acoustic field (homogeneous saturation), can be modeled with dynamical equations~\cite{Lecoeuche1996a}, which makes it is difficult to devise an accurate estimate for $ P_{\text{max}} $.
Nonetheless, a coarse scaling law can be established in relation with the gain at the mode adjacent to the lasing mode, knowing Brillouin gain is Lorentzian:
\begin{equation}\label{eq:PmaxRatio}
	\dfrac{P_{\text{max}}}{P_{th}^{j}} \approx 1 + \left[ 2 \dfrac{D_1}{\Gamma_B} \right]^2 , \;
	P_{th}^{j}
	\approx \dfrac{\Aeff}{g_B} \dfrac{D_1 \, \left[ 1 - \eta \,(1-\theta) \right]}{2\pi c}
\end{equation}
where $ P_{th}^{j} $ is the lasing threshold~\cite{Randoux1995} (assuming the gain is on resonance, $ \delta\Omega_j = 0 $).
Note that the above expression typically underestimates $ P_{\text{max}} $, as we neglect mode competition and hysteresis effects, which tend to extend the single mode operation of an initially lasing mode.

\Cref{fig:Concept:NoiseLength} maps the relations in \cref{eq:PmaxRatio}, to illustrate our discussion.
A longer cavity exhibits pure multimode lasing at lower power $ P_{\text{max}} $.
In our study, we measured $ P_{\text{max}} \sim \SI{220}{\milli\watt} $ in the \SI{60}{\meter}-long cavity B, and $ P_{\text{max}} \sim \SI{470}{\milli\watt} $ in the \SI{10}{\meter}-long cavity A.
Beyond a critical length, there is no single-mode Stokes emission.
On the other hand, short cavities are interesting to preserve the high-coherence, single-mode operation over a wider range of power.
Such cavities also enable access to a wider span of detuning and hence a larger variety of comb states, such as the soliton state in the anomalous dispersion regime,  which appears at a relatively large detuning~\cite{Leo2010}.
However, shorter cavities exhibit a higher lasing threshold and a lower accumulated nonlinearity.
Indeed, in fiber cavities, the loss per roundtrip is dominated by lumped elements such as the circulator.
Hence, the finesse remains limited and the threshold increases with the FSR.
Microresonators present high finesse and nonlinearity, thus circumvent most of these shortcomings and allow soliton-based combs~\cite{Bai2020,Do2021}, but require resonant pumping and precise matching of the cavity's FSR with the Stokes shift, which complicates detuning control.
In conclusion, while the generation of high repetition rate Kerr combs in a fiber Brillouin laser cavity has multiple potential advantages in terms of reconfigurability, pump coupling efficiency and coherence, one must be aware of the limitations in these systems, especially in view of the targeted applications.
Further analysis of the influence of other cavity parameters will be the subject of another publication.
Our observations help to understand the comb states reported in refs.~\cite{Li2017c,Huang2019a}, where \SIrange{300}{500}{\meter}-long fibers are used.
It is highly unlikely that the regime observed in ref.~\cite{Huang2019a} corresponds to a stable soliton state as claimed.
Although more than two pumps are used simultaneously in this work, the length of the fiber used is not compatible with stable laser operation, nor with access to the detuning range of soliton existence.
Interestingly, we can extend the conclusions of our analysis to a greater number of pumps.
Preserving coherent operation in a multi-pump scenario likely puts even more stringent constrains on synchronizing the pumps spacing with the FSR.
Moreover, the variation of the Brillouin phonon frequency with wavelength, would intrinsically induce unequal pump detunings, making it harder to maintain coherent operation.

\begin{acknowledgments}
	The authors acknowledge support from French program ``Investments for the Future'' operated by the National Research Agency (EIPHI Graduate School, contract ANR-17-EURE-0002), and from Région Bourgogne Franche-Comté and European Regional Development Fund.
\end{acknowledgments}

\bibliography{library}

\begin{thebibliography}{26}%
\makeatletter
\providecommand \@ifxundefined [1]{%
 \@ifx{#1\undefined}
}%
\providecommand \@ifnum [1]{%
 \ifnum #1\expandafter \@firstoftwo
 \else \expandafter \@secondoftwo
 \fi
}%
\providecommand \@ifx [1]{%
 \ifx #1\expandafter \@firstoftwo
 \else \expandafter \@secondoftwo
 \fi
}%
\providecommand \natexlab [1]{#1}%
\providecommand \enquote  [1]{``#1''}%
\providecommand \bibnamefont  [1]{#1}%
\providecommand \bibfnamefont [1]{#1}%
\providecommand \citenamefont [1]{#1}%
\providecommand \href@noop [0]{\@secondoftwo}%
\providecommand \href [0]{\begingroup \@sanitize@url \@href}%
\providecommand \@href[1]{\@@startlink{#1}\@@href}%
\providecommand \@@href[1]{\endgroup#1\@@endlink}%
\providecommand \@sanitize@url [0]{\catcode `\\12\catcode `\$12\catcode
  `\&12\catcode `\#12\catcode `\^12\catcode `\_12\catcode `\%12\relax}%
\providecommand \@@startlink[1]{}%
\providecommand \@@endlink[0]{}%
\providecommand \url  [0]{\begingroup\@sanitize@url \@url }%
\providecommand \@url [1]{\endgroup\@href {#1}{\urlprefix }}%
\providecommand \urlprefix  [0]{URL }%
\providecommand \Eprint [0]{\href }%
\providecommand \doibase [0]{https://doi.org/}%
\providecommand \selectlanguage [0]{\@gobble}%
\providecommand \bibinfo  [0]{\@secondoftwo}%
\providecommand \bibfield  [0]{\@secondoftwo}%
\providecommand \translation [1]{[#1]}%
\providecommand \BibitemOpen [0]{}%
\providecommand \bibitemStop [0]{}%
\providecommand \bibitemNoStop [0]{.\EOS\space}%
\providecommand \EOS [0]{\spacefactor3000\relax}%
\providecommand \BibitemShut  [1]{\csname bibitem#1\endcsname}%
\let\auto@bib@innerbib\@empty
\bibitem [{\citenamefont {Diddams}\ \emph {et~al.}(2020)\citenamefont
  {Diddams}, \citenamefont {Vahala},\ and\ \citenamefont {Udem}}]{Diddams2020}%
  \BibitemOpen
  \bibfield  {author} {\bibinfo {author} {\bibfnamefont {S.~A.}\ \bibnamefont
  {Diddams}}, \bibinfo {author} {\bibfnamefont {K.}~\bibnamefont {Vahala}},\
  and\ \bibinfo {author} {\bibfnamefont {T.}~\bibnamefont {Udem}},\ }\bibfield
  {title} {\bibinfo {title} {Optical frequency combs: {{Coherently}} uniting
  the electromagnetic spectrum},\ }\href
  {https://doi.org/10.1126/science.aay3676} {\bibfield  {journal} {\bibinfo
  {journal} {Science}\ }\textbf {\bibinfo {volume} {369}},\ \bibinfo {pages}
  {eaay3676} (\bibinfo {year} {2020})}\BibitemShut {NoStop}%
\bibitem [{\citenamefont {Picqu{\'e}}\ and\ \citenamefont
  {H{\"a}nsch}(2019)}]{Picque2019}%
  \BibitemOpen
  \bibfield  {author} {\bibinfo {author} {\bibfnamefont {N.}~\bibnamefont
  {Picqu{\'e}}}\ and\ \bibinfo {author} {\bibfnamefont {T.~W.}\ \bibnamefont
  {H{\"a}nsch}},\ }\bibfield  {title} {\bibinfo {title} {Frequency comb
  spectroscopy},\ }\href {https://doi.org/10.1038/s41566-018-0347-5} {\bibfield
   {journal} {\bibinfo  {journal} {Nature Photonics}\ }\textbf {\bibinfo
  {volume} {13}},\ \bibinfo {pages} {146} (\bibinfo {year} {2019})}\BibitemShut
  {NoStop}%
\bibitem [{\citenamefont {Li}\ \emph {et~al.}(2017)\citenamefont {Li},
  \citenamefont {Jia}, \citenamefont {Li}, \citenamefont {Yang}, \citenamefont
  {Xiao}, \citenamefont {Chen}, \citenamefont {Qin}, \citenamefont {Huang},\
  and\ \citenamefont {Qin}}]{Li2017c}%
  \BibitemOpen
  \bibfield  {author} {\bibinfo {author} {\bibfnamefont {Q.}~\bibnamefont
  {Li}}, \bibinfo {author} {\bibfnamefont {Z.-x.}\ \bibnamefont {Jia}},
  \bibinfo {author} {\bibfnamefont {Z.-r.}\ \bibnamefont {Li}}, \bibinfo
  {author} {\bibfnamefont {Y.-d.}\ \bibnamefont {Yang}}, \bibinfo {author}
  {\bibfnamefont {J.-l.}\ \bibnamefont {Xiao}}, \bibinfo {author}
  {\bibfnamefont {S.-w.}\ \bibnamefont {Chen}}, \bibinfo {author}
  {\bibfnamefont {G.-s.}\ \bibnamefont {Qin}}, \bibinfo {author} {\bibfnamefont
  {Y.-z.}\ \bibnamefont {Huang}},\ and\ \bibinfo {author} {\bibfnamefont
  {W.-p.}\ \bibnamefont {Qin}},\ }\bibfield  {title} {\bibinfo {title} {Optical
  frequency combs generated by four-wave mixing in a dual wavelength
  {{Brillouin}} laser cavity},\ }\href {https://doi.org/10.1063/1.4994861}
  {\bibfield  {journal} {\bibinfo  {journal} {AIP Advances}\ }\textbf {\bibinfo
  {volume} {7}},\ \bibinfo {pages} {075215} (\bibinfo {year}
  {2017})}\BibitemShut {NoStop}%
\bibitem [{\citenamefont {Kippenberg}\ \emph {et~al.}(2018)\citenamefont
  {Kippenberg}, \citenamefont {Gaeta}, \citenamefont {Lipson},\ and\
  \citenamefont {Gorodetsky}}]{Kippenberg2018}%
  \BibitemOpen
  \bibfield  {author} {\bibinfo {author} {\bibfnamefont {T.~J.}\ \bibnamefont
  {Kippenberg}}, \bibinfo {author} {\bibfnamefont {A.~L.}\ \bibnamefont
  {Gaeta}}, \bibinfo {author} {\bibfnamefont {M.}~\bibnamefont {Lipson}},\ and\
  \bibinfo {author} {\bibfnamefont {M.~L.}\ \bibnamefont {Gorodetsky}},\
  }\bibfield  {title} {\bibinfo {title} {Dissipative {{Kerr}} solitons in
  optical microresonators},\ }\href {https://doi.org/10.1126/science.aan8083}
  {\bibfield  {journal} {\bibinfo  {journal} {Science}\ }\textbf {\bibinfo
  {volume} {361}},\ \bibinfo {pages} {eaan8083} (\bibinfo {year}
  {2018})}\BibitemShut {NoStop}%
\bibitem [{\citenamefont {Huang}\ \emph {et~al.}(2019)\citenamefont {Huang},
  \citenamefont {Li}, \citenamefont {Han}, \citenamefont {Jia}, \citenamefont
  {Yu}, \citenamefont {Yang}, \citenamefont {Xiao}, \citenamefont {Wu},
  \citenamefont {Zhang}, \citenamefont {Huang}, \citenamefont {Qin},\ and\
  \citenamefont {Qin}}]{Huang2019a}%
  \BibitemOpen
  \bibfield  {author} {\bibinfo {author} {\bibfnamefont {Y.}~\bibnamefont
  {Huang}}, \bibinfo {author} {\bibfnamefont {Q.}~\bibnamefont {Li}}, \bibinfo
  {author} {\bibfnamefont {J.}~\bibnamefont {Han}}, \bibinfo {author}
  {\bibfnamefont {Z.}~\bibnamefont {Jia}}, \bibinfo {author} {\bibfnamefont
  {Y.}~\bibnamefont {Yu}}, \bibinfo {author} {\bibfnamefont {Y.}~\bibnamefont
  {Yang}}, \bibinfo {author} {\bibfnamefont {J.}~\bibnamefont {Xiao}}, \bibinfo
  {author} {\bibfnamefont {J.}~\bibnamefont {Wu}}, \bibinfo {author}
  {\bibfnamefont {D.}~\bibnamefont {Zhang}}, \bibinfo {author} {\bibfnamefont
  {Y.}~\bibnamefont {Huang}}, \bibinfo {author} {\bibfnamefont
  {W.}~\bibnamefont {Qin}},\ and\ \bibinfo {author} {\bibfnamefont
  {G.}~\bibnamefont {Qin}},\ }\bibfield  {title} {\bibinfo {title} {Temporal
  soliton and optical frequency comb generation in a {{Brillouin}} laser
  cavity},\ }\href {https://doi.org/10.1364/OPTICA.6.001491} {\bibfield
  {journal} {\bibinfo  {journal} {Optica}\ }\textbf {\bibinfo {volume} {6}},\
  \bibinfo {pages} {1491} (\bibinfo {year} {2019})}\BibitemShut {NoStop}%
\bibitem [{\citenamefont {Bai}\ \emph {et~al.}(2021)\citenamefont {Bai},
  \citenamefont {Zhang}, \citenamefont {Shi}, \citenamefont {Ding},
  \citenamefont {Qin}, \citenamefont {Xie}, \citenamefont {Jiang},\ and\
  \citenamefont {Xiao}}]{Bai2020}%
  \BibitemOpen
  \bibfield  {author} {\bibinfo {author} {\bibfnamefont {Y.}~\bibnamefont
  {Bai}}, \bibinfo {author} {\bibfnamefont {M.}~\bibnamefont {Zhang}}, \bibinfo
  {author} {\bibfnamefont {Q.}~\bibnamefont {Shi}}, \bibinfo {author}
  {\bibfnamefont {S.}~\bibnamefont {Ding}}, \bibinfo {author} {\bibfnamefont
  {Y.}~\bibnamefont {Qin}}, \bibinfo {author} {\bibfnamefont {Z.}~\bibnamefont
  {Xie}}, \bibinfo {author} {\bibfnamefont {X.}~\bibnamefont {Jiang}},\ and\
  \bibinfo {author} {\bibfnamefont {M.}~\bibnamefont {Xiao}},\ }\bibfield
  {title} {\bibinfo {title} {Brillouin-{{Kerr Soliton Frequency Combs}} in an
  {{Optical Microresonator}}},\ }\href
  {https://doi.org/10.1103/PhysRevLett.126.063901} {\bibfield  {journal}
  {\bibinfo  {journal} {Physical Review Letters}\ }\textbf {\bibinfo {volume}
  {126}},\ \bibinfo {pages} {063901} (\bibinfo {year} {2021})}\BibitemShut
  {NoStop}%
\bibitem [{\citenamefont {Do}\ \emph {et~al.}(2021)\citenamefont {Do},
  \citenamefont {Kim}, \citenamefont {Jeong}, \citenamefont {Suk},
  \citenamefont {Kwon}, \citenamefont {Kim}, \citenamefont {Lee},\ and\
  \citenamefont {Lee}}]{Do2021}%
  \BibitemOpen
  \bibfield  {author} {\bibinfo {author} {\bibfnamefont {I.~H.}\ \bibnamefont
  {Do}}, \bibinfo {author} {\bibfnamefont {D.}~\bibnamefont {Kim}}, \bibinfo
  {author} {\bibfnamefont {D.}~\bibnamefont {Jeong}}, \bibinfo {author}
  {\bibfnamefont {D.}~\bibnamefont {Suk}}, \bibinfo {author} {\bibfnamefont
  {D.}~\bibnamefont {Kwon}}, \bibinfo {author} {\bibfnamefont {J.}~\bibnamefont
  {Kim}}, \bibinfo {author} {\bibfnamefont {J.~H.}\ \bibnamefont {Lee}},\ and\
  \bibinfo {author} {\bibfnamefont {H.}~\bibnamefont {Lee}},\ }\bibfield
  {title} {\bibinfo {title} {Self-stabilized soliton generation in a
  microresonator through mode-pulled {{Brillouin}} lasing},\ }\href
  {https://doi.org/10.1364/OL.419137} {\bibfield  {journal} {\bibinfo
  {journal} {Optics Letters}\ }\textbf {\bibinfo {volume} {46}},\ \bibinfo
  {pages} {1772} (\bibinfo {year} {2021})}\BibitemShut {NoStop}%
\bibitem [{\citenamefont {Shirazi}\ \emph {et~al.}(2007)\citenamefont
  {Shirazi}, \citenamefont {Harun}, \citenamefont {Thambiratnam}, \citenamefont
  {Biglary},\ and\ \citenamefont {Ahmad}}]{Shirazi2007}%
  \BibitemOpen
  \bibfield  {author} {\bibinfo {author} {\bibfnamefont {M.~R.}\ \bibnamefont
  {Shirazi}}, \bibinfo {author} {\bibfnamefont {S.~W.}\ \bibnamefont {Harun}},
  \bibinfo {author} {\bibfnamefont {K.}~\bibnamefont {Thambiratnam}}, \bibinfo
  {author} {\bibfnamefont {M.}~\bibnamefont {Biglary}},\ and\ \bibinfo {author}
  {\bibfnamefont {H.}~\bibnamefont {Ahmad}},\ }\bibfield  {title} {\bibinfo
  {title} {New {{Brillouin}} fiber laser configuration with high output
  power},\ }\href {https://doi.org/10.1002/mop.22805} {\bibfield  {journal}
  {\bibinfo  {journal} {Microwave and Optical Technology Letters}\ }\textbf
  {\bibinfo {volume} {49}},\ \bibinfo {pages} {2656} (\bibinfo {year}
  {2007})}\BibitemShut {NoStop}%
\bibitem [{\citenamefont {Danion}\ \emph {et~al.}(2016)\citenamefont {Danion},
  \citenamefont {Frein}, \citenamefont {Bacquet}, \citenamefont {Pillet},
  \citenamefont {Molin}, \citenamefont {Morvan}, \citenamefont {Ducournau},
  \citenamefont {Vallet}, \citenamefont {Szriftgiser},\ and\ \citenamefont
  {Alouini}}]{Danion2016}%
  \BibitemOpen
  \bibfield  {author} {\bibinfo {author} {\bibfnamefont {G.}~\bibnamefont
  {Danion}}, \bibinfo {author} {\bibfnamefont {L.}~\bibnamefont {Frein}},
  \bibinfo {author} {\bibfnamefont {D.}~\bibnamefont {Bacquet}}, \bibinfo
  {author} {\bibfnamefont {G.}~\bibnamefont {Pillet}}, \bibinfo {author}
  {\bibfnamefont {S.}~\bibnamefont {Molin}}, \bibinfo {author} {\bibfnamefont
  {L.}~\bibnamefont {Morvan}}, \bibinfo {author} {\bibfnamefont
  {G.}~\bibnamefont {Ducournau}}, \bibinfo {author} {\bibfnamefont
  {M.}~\bibnamefont {Vallet}}, \bibinfo {author} {\bibfnamefont
  {P.}~\bibnamefont {Szriftgiser}},\ and\ \bibinfo {author} {\bibfnamefont
  {M.}~\bibnamefont {Alouini}},\ }\bibfield  {title} {\bibinfo {title}
  {Mode-hopping suppression in long {{Brillouin}} fiber laser with non-resonant
  pumping},\ }\href {https://doi.org/10.1364/OL.41.002362} {\bibfield
  {journal} {\bibinfo  {journal} {Optics Letters}\ }\textbf {\bibinfo {volume}
  {41}},\ \bibinfo {pages} {2362} (\bibinfo {year} {2016})}\BibitemShut
  {NoStop}%
\bibitem [{\citenamefont {Gundavarapu}\ \emph {et~al.}(2019)\citenamefont
  {Gundavarapu}, \citenamefont {Brodnik}, \citenamefont {Puckett},
  \citenamefont {Huffman}, \citenamefont {Bose}, \citenamefont {Behunin},
  \citenamefont {Wu}, \citenamefont {Qiu}, \citenamefont {Pinho}, \citenamefont
  {Chauhan}, \citenamefont {Nohava}, \citenamefont {Rakich}, \citenamefont
  {Nelson}, \citenamefont {Salit},\ and\ \citenamefont
  {Blumenthal}}]{Gundavarapu2019}%
  \BibitemOpen
  \bibfield  {author} {\bibinfo {author} {\bibfnamefont {S.}~\bibnamefont
  {Gundavarapu}}, \bibinfo {author} {\bibfnamefont {G.~M.}\ \bibnamefont
  {Brodnik}}, \bibinfo {author} {\bibfnamefont {M.}~\bibnamefont {Puckett}},
  \bibinfo {author} {\bibfnamefont {T.}~\bibnamefont {Huffman}}, \bibinfo
  {author} {\bibfnamefont {D.}~\bibnamefont {Bose}}, \bibinfo {author}
  {\bibfnamefont {R.}~\bibnamefont {Behunin}}, \bibinfo {author} {\bibfnamefont
  {J.}~\bibnamefont {Wu}}, \bibinfo {author} {\bibfnamefont {T.}~\bibnamefont
  {Qiu}}, \bibinfo {author} {\bibfnamefont {C.}~\bibnamefont {Pinho}}, \bibinfo
  {author} {\bibfnamefont {N.}~\bibnamefont {Chauhan}}, \bibinfo {author}
  {\bibfnamefont {J.}~\bibnamefont {Nohava}}, \bibinfo {author} {\bibfnamefont
  {P.~T.}\ \bibnamefont {Rakich}}, \bibinfo {author} {\bibfnamefont {K.~D.}\
  \bibnamefont {Nelson}}, \bibinfo {author} {\bibfnamefont {M.}~\bibnamefont
  {Salit}},\ and\ \bibinfo {author} {\bibfnamefont {D.~J.}\ \bibnamefont
  {Blumenthal}},\ }\bibfield  {title} {\bibinfo {title} {Sub-hertz fundamental
  linewidth photonic integrated {{Brillouin}} laser},\ }\href
  {https://doi.org/10.1038/s41566-018-0313-2} {\bibfield  {journal} {\bibinfo
  {journal} {Nature Photonics}\ }\textbf {\bibinfo {volume} {13}},\ \bibinfo
  {pages} {60} (\bibinfo {year} {2019})}\BibitemShut {NoStop}%
\bibitem [{\citenamefont {Debut}\ \emph {et~al.}(2000)\citenamefont {Debut},
  \citenamefont {Randoux},\ and\ \citenamefont {Zemmouri}}]{Debut2000}%
  \BibitemOpen
  \bibfield  {author} {\bibinfo {author} {\bibfnamefont {A.}~\bibnamefont
  {Debut}}, \bibinfo {author} {\bibfnamefont {S.}~\bibnamefont {Randoux}},\
  and\ \bibinfo {author} {\bibfnamefont {J.}~\bibnamefont {Zemmouri}},\
  }\bibfield  {title} {\bibinfo {title} {Linewidth narrowing in {{Brillouin}}
  lasers: {{Theoretical}} analysis},\ }\href
  {https://doi.org/10.1103/PhysRevA.62.023803} {\bibfield  {journal} {\bibinfo
  {journal} {Physical Review A}\ }\textbf {\bibinfo {volume} {62}},\ \bibinfo
  {pages} {023803} (\bibinfo {year} {2000})}\BibitemShut {NoStop}%
\bibitem [{\citenamefont {Hansson}\ and\ \citenamefont
  {Wabnitz}(2014)}]{Hansson2014a}%
  \BibitemOpen
  \bibfield  {author} {\bibinfo {author} {\bibfnamefont {T.}~\bibnamefont
  {Hansson}}\ and\ \bibinfo {author} {\bibfnamefont {S.}~\bibnamefont
  {Wabnitz}},\ }\bibfield  {title} {\bibinfo {title} {Bichromatically pumped
  microresonator frequency combs},\ }\href
  {https://doi.org/10.1103/PhysRevA.90.013811} {\bibfield  {journal} {\bibinfo
  {journal} {Physical Review A}\ }\textbf {\bibinfo {volume} {90}},\ \bibinfo
  {pages} {013811} (\bibinfo {year} {2014})}\BibitemShut {NoStop}%
\bibitem [{\citenamefont {Ikeda}(1979)}]{Ikeda1979}%
  \BibitemOpen
  \bibfield  {author} {\bibinfo {author} {\bibfnamefont {K.}~\bibnamefont
  {Ikeda}},\ }\bibfield  {title} {\bibinfo {title} {Multiple-valued stationary
  state and its instability of the transmitted light by a ring cavity system},\
  }\href {https://doi.org/10.1016/0030-4018(79)90090-7} {\bibfield  {journal}
  {\bibinfo  {journal} {Optics Communications}\ }\textbf {\bibinfo {volume}
  {30}},\ \bibinfo {pages} {257} (\bibinfo {year} {1979})}\BibitemShut
  {NoStop}%
\bibitem [{\citenamefont {Kierzenka}\ and\ \citenamefont
  {Shampine}(2001)}]{Kierzenka2001}%
  \BibitemOpen
  \bibfield  {author} {\bibinfo {author} {\bibfnamefont {J.}~\bibnamefont
  {Kierzenka}}\ and\ \bibinfo {author} {\bibfnamefont {L.~F.}\ \bibnamefont
  {Shampine}},\ }\bibfield  {title} {\bibinfo {title} {A {{BVP}} solver based
  on residual control and the {{Maltab PSE}}},\ }\href
  {https://doi.org/10.1145/502800.502801} {\bibfield  {journal} {\bibinfo
  {journal} {ACM Transactions on Mathematical Software}\ }\textbf {\bibinfo
  {volume} {27}},\ \bibinfo {pages} {299} (\bibinfo {year} {2001})}\BibitemShut
  {NoStop}%
\bibitem [{\citenamefont {Deroh}\ \emph {et~al.}(2018)\citenamefont {Deroh},
  \citenamefont {Kibler}, \citenamefont {Maillotte}, \citenamefont
  {Sylvestre},\ and\ \citenamefont {Beugnot}}]{DerohLargeBrillouin2018}%
  \BibitemOpen
  \bibfield  {author} {\bibinfo {author} {\bibfnamefont {M.}~\bibnamefont
  {Deroh}}, \bibinfo {author} {\bibfnamefont {B.}~\bibnamefont {Kibler}},
  \bibinfo {author} {\bibfnamefont {H.}~\bibnamefont {Maillotte}}, \bibinfo
  {author} {\bibfnamefont {T.}~\bibnamefont {Sylvestre}},\ and\ \bibinfo
  {author} {\bibfnamefont {J.-C.}\ \bibnamefont {Beugnot}},\ }\bibfield
  {title} {\bibinfo {title} {Large {{Brillouin}} gain in {{Germania-doped}}
  core optical fibers up to a 98 mol\% doping level},\ }\href
  {https://doi.org/10.1364/OL.43.004005} {\bibfield  {journal} {\bibinfo
  {journal} {Optics Letters}\ }\textbf {\bibinfo {volume} {43}},\ \bibinfo
  {pages} {4005} (\bibinfo {year} {2018})}\BibitemShut {NoStop}%
\bibitem [{\citenamefont {Thorpe}\ \emph {et~al.}(2008)\citenamefont {Thorpe},
  \citenamefont {Numata},\ and\ \citenamefont {Livas}}]{Thorpe2008b}%
  \BibitemOpen
  \bibfield  {author} {\bibinfo {author} {\bibfnamefont {J.~I.}\ \bibnamefont
  {Thorpe}}, \bibinfo {author} {\bibfnamefont {K.}~\bibnamefont {Numata}},\
  and\ \bibinfo {author} {\bibfnamefont {J.}~\bibnamefont {Livas}},\ }\bibfield
   {title} {\bibinfo {title} {Laser frequency stabilization and control through
  offset sideband locking to optical cavities},\ }\href
  {https://doi.org/10.1364/OE.16.015980} {\bibfield  {journal} {\bibinfo
  {journal} {Optics Express}\ }\textbf {\bibinfo {volume} {16}},\ \bibinfo
  {pages} {15980} (\bibinfo {year} {2008})}\BibitemShut {NoStop}%
\bibitem [{\citenamefont {Nicati}\ \emph {et~al.}(1994)\citenamefont {Nicati},
  \citenamefont {Toyama}, \citenamefont {Huang},\ and\ \citenamefont
  {Shaw}}]{Nicati1994a}%
  \BibitemOpen
  \bibfield  {author} {\bibinfo {author} {\bibfnamefont {P.-A.}\ \bibnamefont
  {Nicati}}, \bibinfo {author} {\bibfnamefont {K.}~\bibnamefont {Toyama}},
  \bibinfo {author} {\bibfnamefont {S.}~\bibnamefont {Huang}},\ and\ \bibinfo
  {author} {\bibfnamefont {H.}~\bibnamefont {Shaw}},\ }\bibfield  {title}
  {\bibinfo {title} {Frequency pulling in a {{Brillouin}} fiber ring laser},\
  }\href {https://doi.org/10.1109/68.311459} {\bibfield  {journal} {\bibinfo
  {journal} {IEEE Photonics Technology Letters}\ }\textbf {\bibinfo {volume}
  {6}},\ \bibinfo {pages} {801} (\bibinfo {year} {1994})}\BibitemShut {NoStop}%
\bibitem [{\citenamefont {Li}\ \emph {et~al.}(2012)\citenamefont {Li},
  \citenamefont {Lee}, \citenamefont {Chen},\ and\ \citenamefont
  {Vahala}}]{Li2012a}%
  \BibitemOpen
  \bibfield  {author} {\bibinfo {author} {\bibfnamefont {J.}~\bibnamefont
  {Li}}, \bibinfo {author} {\bibfnamefont {H.}~\bibnamefont {Lee}}, \bibinfo
  {author} {\bibfnamefont {T.}~\bibnamefont {Chen}},\ and\ \bibinfo {author}
  {\bibfnamefont {K.~J.}\ \bibnamefont {Vahala}},\ }\bibfield  {title}
  {\bibinfo {title} {Characterization of a high coherence, {{Brillouin}}
  microcavity laser on silicon},\ }\href {https://doi.org/10.1364/OE.20.020170}
  {\bibfield  {journal} {\bibinfo  {journal} {Optics Express}\ }\textbf
  {\bibinfo {volume} {20}},\ \bibinfo {pages} {20170} (\bibinfo {year}
  {2012})}\BibitemShut {NoStop}%
\bibitem [{\citenamefont {Lecoeuche}\ \emph
  {et~al.}(1996{\natexlab{a}})\citenamefont {Lecoeuche}, \citenamefont
  {Randoux}, \citenamefont {S{\'e}gard},\ and\ \citenamefont
  {Zemmouri}}]{Lecoeuche1996}%
  \BibitemOpen
  \bibfield  {author} {\bibinfo {author} {\bibfnamefont {V.}~\bibnamefont
  {Lecoeuche}}, \bibinfo {author} {\bibfnamefont {S.}~\bibnamefont {Randoux}},
  \bibinfo {author} {\bibfnamefont {B.}~\bibnamefont {S{\'e}gard}},\ and\
  \bibinfo {author} {\bibfnamefont {J.}~\bibnamefont {Zemmouri}},\ }\bibfield
  {title} {\bibinfo {title} {Dynamics of stimulated {{Brillouin}} scattering
  with feedback},\ }\href {https://doi.org/10.1088/1355-5111/8/6/003}
  {\bibfield  {journal} {\bibinfo  {journal} {Quantum and Semiclassical Optics:
  Journal of the European Optical Society Part B}\ }\textbf {\bibinfo {volume}
  {8}},\ \bibinfo {pages} {1109} (\bibinfo {year}
  {1996}{\natexlab{a}})}\BibitemShut {NoStop}%
\bibitem [{\citenamefont {Strekalov}\ and\ \citenamefont
  {Yu}(2009)}]{Strekalov2009}%
  \BibitemOpen
  \bibfield  {author} {\bibinfo {author} {\bibfnamefont {D.~V.}\ \bibnamefont
  {Strekalov}}\ and\ \bibinfo {author} {\bibfnamefont {N.}~\bibnamefont {Yu}},\
  }\bibfield  {title} {\bibinfo {title} {Generation of optical combs in a
  whispering gallery mode resonator from a bichromatic pump},\ }\href
  {https://doi.org/10.1103/PhysRevA.79.041805} {\bibfield  {journal} {\bibinfo
  {journal} {Physical Review A}\ }\textbf {\bibinfo {volume} {79}},\ \bibinfo
  {pages} {041805} (\bibinfo {year} {2009})}\BibitemShut {NoStop}%
\bibitem [{\citenamefont {Garbin}\ \emph {et~al.}(2017)\citenamefont {Garbin},
  \citenamefont {Wang}, \citenamefont {Murdoch}, \citenamefont {Oppo},
  \citenamefont {Coen},\ and\ \citenamefont {Erkintalo}}]{Garbin2017}%
  \BibitemOpen
  \bibfield  {author} {\bibinfo {author} {\bibfnamefont {B.}~\bibnamefont
  {Garbin}}, \bibinfo {author} {\bibfnamefont {Y.}~\bibnamefont {Wang}},
  \bibinfo {author} {\bibfnamefont {S.~G.}\ \bibnamefont {Murdoch}}, \bibinfo
  {author} {\bibfnamefont {G.-L.}\ \bibnamefont {Oppo}}, \bibinfo {author}
  {\bibfnamefont {S.}~\bibnamefont {Coen}},\ and\ \bibinfo {author}
  {\bibfnamefont {M.}~\bibnamefont {Erkintalo}},\ }\bibfield  {title} {\bibinfo
  {title} {Experimental and numerical investigations of switching wave dynamics
  in a normally dispersive fibre ring resonator},\ }\href
  {https://doi.org/10.1140/epjd/e2017-80133-7} {\bibfield  {journal} {\bibinfo
  {journal} {The European Physical Journal D}\ }\textbf {\bibinfo {volume}
  {71}},\ \bibinfo {pages} {240} (\bibinfo {year} {2017})}\BibitemShut
  {NoStop}%
\bibitem [{\citenamefont {Lucas}\ \emph {et~al.}(2017)\citenamefont {Lucas},
  \citenamefont {Guo}, \citenamefont {Jost}, \citenamefont {Karpov},\ and\
  \citenamefont {Kippenberg}}]{Lucas2017}%
  \BibitemOpen
  \bibfield  {author} {\bibinfo {author} {\bibfnamefont {E.}~\bibnamefont
  {Lucas}}, \bibinfo {author} {\bibfnamefont {H.}~\bibnamefont {Guo}}, \bibinfo
  {author} {\bibfnamefont {J.~D.}\ \bibnamefont {Jost}}, \bibinfo {author}
  {\bibfnamefont {M.}~\bibnamefont {Karpov}},\ and\ \bibinfo {author}
  {\bibfnamefont {T.~J.}\ \bibnamefont {Kippenberg}},\ }\bibfield  {title}
  {\bibinfo {title} {Detuning-dependent properties and dispersion-induced
  instabilities of temporal dissipative {{Kerr}} solitons in optical
  microresonators},\ }\href {https://doi.org/10.1103/PhysRevA.95.043822}
  {\bibfield  {journal} {\bibinfo  {journal} {Physical Review A}\ }\textbf
  {\bibinfo {volume} {95}},\ \bibinfo {pages} {043822} (\bibinfo {year}
  {2017})}\BibitemShut {NoStop}%
\bibitem [{\citenamefont {Obrzud}\ \emph {et~al.}(2017)\citenamefont {Obrzud},
  \citenamefont {Lecomte},\ and\ \citenamefont {Herr}}]{Obrzud2017}%
  \BibitemOpen
  \bibfield  {author} {\bibinfo {author} {\bibfnamefont {E.}~\bibnamefont
  {Obrzud}}, \bibinfo {author} {\bibfnamefont {S.}~\bibnamefont {Lecomte}},\
  and\ \bibinfo {author} {\bibfnamefont {T.}~\bibnamefont {Herr}},\ }\bibfield
  {title} {\bibinfo {title} {Temporal solitons in microresonators driven by
  optical pulses},\ }\href {https://doi.org/10.1038/nphoton.2017.140}
  {\bibfield  {journal} {\bibinfo  {journal} {Nature Photonics}\ }\textbf
  {\bibinfo {volume} {11}},\ \bibinfo {pages} {600} (\bibinfo {year}
  {2017})}\BibitemShut {NoStop}%
\bibitem [{\citenamefont {Randoux}\ \emph {et~al.}(1995)\citenamefont
  {Randoux}, \citenamefont {Lecoeuche}, \citenamefont {S{\'e}gard},\ and\
  \citenamefont {Zemmouri}}]{Randoux1995}%
  \BibitemOpen
  \bibfield  {author} {\bibinfo {author} {\bibfnamefont {S.}~\bibnamefont
  {Randoux}}, \bibinfo {author} {\bibfnamefont {V.}~\bibnamefont {Lecoeuche}},
  \bibinfo {author} {\bibfnamefont {B.}~\bibnamefont {S{\'e}gard}},\ and\
  \bibinfo {author} {\bibfnamefont {J.}~\bibnamefont {Zemmouri}},\ }\bibfield
  {title} {\bibinfo {title} {Dynamical analysis of {{Brillouin}} fiber lasers:
  {{An}} experimental approach},\ }\href
  {https://doi.org/10.1103/PhysRevA.51.R4345} {\bibfield  {journal} {\bibinfo
  {journal} {Physical Review A}\ }\textbf {\bibinfo {volume} {51}},\ \bibinfo
  {pages} {R4345} (\bibinfo {year} {1995})}\BibitemShut {NoStop}%
\bibitem [{\citenamefont {Lecoeuche}\ \emph
  {et~al.}(1996{\natexlab{b}})\citenamefont {Lecoeuche}, \citenamefont
  {Randoux}, \citenamefont {S{\'e}gard},\ and\ \citenamefont
  {Zemmouri}}]{Lecoeuche1996a}%
  \BibitemOpen
  \bibfield  {author} {\bibinfo {author} {\bibfnamefont {V.}~\bibnamefont
  {Lecoeuche}}, \bibinfo {author} {\bibfnamefont {S.}~\bibnamefont {Randoux}},
  \bibinfo {author} {\bibfnamefont {B.}~\bibnamefont {S{\'e}gard}},\ and\
  \bibinfo {author} {\bibfnamefont {J.}~\bibnamefont {Zemmouri}},\ }\bibfield
  {title} {\bibinfo {title} {Dynamics of a {{Brillouin}} fiber ring laser:
  {{Off-resonant}} case},\ }\href {https://doi.org/10.1103/PhysRevA.53.2822}
  {\bibfield  {journal} {\bibinfo  {journal} {Physical Review A}\ }\textbf
  {\bibinfo {volume} {53}},\ \bibinfo {pages} {2822} (\bibinfo {year}
  {1996}{\natexlab{b}})}\BibitemShut {NoStop}%
\bibitem [{\citenamefont {Leo}\ \emph {et~al.}(2010)\citenamefont {Leo},
  \citenamefont {Coen}, \citenamefont {Kockaert}, \citenamefont {Gorza},
  \citenamefont {Emplit},\ and\ \citenamefont {Haelterman}}]{Leo2010}%
  \BibitemOpen
  \bibfield  {author} {\bibinfo {author} {\bibfnamefont {F.}~\bibnamefont
  {Leo}}, \bibinfo {author} {\bibfnamefont {S.}~\bibnamefont {Coen}}, \bibinfo
  {author} {\bibfnamefont {P.}~\bibnamefont {Kockaert}}, \bibinfo {author}
  {\bibfnamefont {S.-P.}\ \bibnamefont {Gorza}}, \bibinfo {author}
  {\bibfnamefont {P.}~\bibnamefont {Emplit}},\ and\ \bibinfo {author}
  {\bibfnamefont {M.}~\bibnamefont {Haelterman}},\ }\bibfield  {title}
  {\bibinfo {title} {Temporal cavity solitons in one-dimensional {{Kerr}} media
  as bits in an all-optical buffer},\ }\href
  {https://doi.org/10.1038/nphoton.2010.120} {\bibfield  {journal} {\bibinfo
  {journal} {Nature Photonics}\ }\textbf {\bibinfo {volume} {4}},\ \bibinfo
  {pages} {471} (\bibinfo {year} {2010})}\BibitemShut {NoStop}%
\end{thebibliography}%

\end{document}